\documentclass[aps,prc,letterpaper,11pt,twoside,tightenlines,nofootinbib,showpacs,preprint,superscriptaddress]{revtex4-1}
\usepackage{float}
\usepackage{epsfig}
\usepackage{subfigure}
\usepackage{amsmath}
\usepackage{rotating}   
\usepackage{xcolor}      
\begin{document}
\arraycolsep1.5pt
\newcommand{\Ima}{\textrm{Im}}
\newcommand{\Rea}{\textrm{Re}}
\newcommand{\mev}{\textrm{ MeV}}
\newcommand{\gev}{\textrm{ GeV}}
\newcommand{\red}[1]{\textcolor{red}{#1}}

\title{Prediction of hidden charm strange molecular baryon
  states with heavy quark spin symmetry}

\author{C. W. Xiao}
\affiliation{School of Physics and Electronics, Central South University, Changsha 410083, China}

\author{J. Nieves}
\affiliation{IFIC, Centro Mixto Universidad de Valencia-CSIC,
Institutos de Investigaci\'on de Paterna, Aptdo. 22085, 46071 Valencia, Spain}

\author{E. Oset}
\affiliation{IFIC, Centro Mixto Universidad de Valencia-CSIC,
Institutos de Investigaci\'on de Paterna, Aptdo. 22085, 46071 Valencia, Spain}
\affiliation{Departamento de F\'{\i}sica Te\'orica, Universidad de Valencia, Spain}
\affiliation{Department of Physics, Guangxi Normal University, Guilin 541004, China}

\date{\today}

\begin{abstract}

  We have studied the meson-baryon $S-$wave interaction in the
  isoscalar hidden-charm strange sector with the coupled-channels,
  $\eta_c \Lambda$, $J/\psi \Lambda$, $\bar{D} \Xi_c$, $\bar{D}_s
  \Lambda_c$, $\bar{D} \Xi_c'$, $\bar{D}^* \Xi_c$, $\bar{D}^*_s
  \Lambda_c$, $\bar{D}^* \Xi_c'$, $\bar{D}^* \Xi_c^*$ in $J^P =
  1/2^-$, $J/\psi \Lambda$, $\bar{D}^* \Xi_c$, $\bar{D}^*_s
  \Lambda_c$, $\bar{D}^* \Xi_c'$ , $\bar{D} \Xi_c^*$, $\bar{D}^*
  \Xi_c^*$ in $3/2^-$ and $\bar{D}^* \Xi_c^*$ in $5/2^-$. We impose
  constraints of heavy quark spin symmetry in the interaction and
  obtain the non vanishing matrix elements from an extension of the
  local hidden gauge approach to the charm sector. The ultraviolet
  divergences are renormalized using the same meson-baryon-loops
  regulator previously employed in the non-strange hidden charm
  sector, where a good reproduction of the properties of the newly
  discovered pentaquark states is obtained. We obtain five states of
  $1/2^-$, four of $3/2^-$ and one of $5/2^-$,
  which could be compared in the near future with forthcoming LHCb
  experiments. The $5/2^-$, three of the $3/2^-$ and another three of
  the $1/2^-$ resonances are originated from isoscalar $\bar
  D^{(*)}\Xi_c^\prime$ and $\bar D^{(*)}\Xi_c^*$ interactions. They
  should be located just few MeV below the corresponding thresholds
  (4446, 4513, 4588 and 4655 MeV), and would be SU(3)-siblings of the
  isospin 1/2 $\bar D^{(*)} \Sigma_c^{(*)}$ quasi-bound states
  previously found, and that provided a robust theoretical description
  of the $P_c(4440)$, $P_c(4457)$ and $P_c(4312)$ LHCb exotic
  states. The another two $1/2^-$ and $3/2^-$ states obtained in this
  work are result of the $\bar D^{(*)}\Xi_c-D^{(*)}_s\Lambda_c$
  coupled-channels isoscalar interaction, are significantly broader
  than the others, with widths of the order of 15 MeV, being $\bar
  D^{(*)}_s\Lambda_c$ the dominant decay channel.

\end{abstract}

\maketitle

\section{Introduction}

The discovery of two pentaquark states of hidden charm in
Refs.~\cite{Aaij:2015tga,Aaij:2015fea} by the LHCb collaboration gave a
boost to hadron physics, providing one example of baryon states that
challenge the standard wisdom of the three quark structure. The Run-2
experiment of the collaboration \cite{Aaij:2019vzc} has added more
precise information where the old narrow state has split into two
differentiated structures and a former small fluctuation has given
rise to a new distinct peak.

Predictions for these hidden charm states in that energy region had
been done before~\cite{Wu:2010jy,Wang:2011rga,Yang:2011wz,Yuan:2012wz,
Wu:2012md,Garcia-Recio:2013gaa, Xiao:2013yca,Uchino:2015uha,Karliner:2015ina}
and, after their discovery, a large number of works were devoted to
explain their possible structure or new reactions where they could be
alternatively obtained. Their detailed discussion and comparison with
other work has been tackled in a series of review papers
\cite{Chen:2016qju,Liu:2019zoy}. The appearance of the new data
\cite{Aaij:2019vzc} has stimulated again a large number of theoretical
papers~\cite{Liu:2019zoy,Chen:2019bip,Liu:2019tjn,He:2019ify,Guo:2019fdo,Chen:2019asm,
  Huang:2019jlf,Ali:2019npk,Shimizu:2019ptd,Xiao:2019mvs,Guo:2019kdc,Cao:2019kst,
  Mutuk:2019snd,Weng:2019ynv,Zhu:2019iwm,Zhang:2019xtu,Wang:2019krd,Eides:2019tgv,
  Wang:2019got,Meng:2019ilv,Gutsche:2019wgu,Cheng:2019obk}.
The vast majority of those works concludes that the new states are of
molecular nature, mostly $\bar{D} \Sigma_c \ (J^P=1/2^-)$, $\bar{D}^*
\Sigma_c \ (J^P=1/2^-,\ 3/2^-)$ with isospin $I=1/2$. In some works
predictions for more states are done, concretely in
Refs.~\cite{Liu:2019tjn} and \cite{Xiao:2019aya} where four more
states of molecular nature involving $\bar{D}^{(*)}$ and
$\Sigma_c^{(*)}$ are reported.

The information provided by the new LHCb experiment
\cite{Aaij:2019vzc} has been very useful to tune the theories for
unknown information. In this sense the masses of the three states
reported in Ref.~\cite{Aaij:2019vzc} have been used in
Ref.~\cite{Xiao:2019aya} to fix the only free parameter of the scheme
of Ref.~\cite{Xiao:2013yca} (a subtraction constant in the regularized meson
baryon loop functions) to agree with the average experimental
mass. With this only experimental input three masses and three widths
are obtained in agreement with experiment in Ref.~\cite{Xiao:2019aya}.

Certainly a theoretical framework is more appreciated when predictions are made prior
to experiment and the latter corroborates the predictions made. In
this sense we intend with the present work to make predictions for
hidden charm molecular states with strangeness, with the hope that
experiments leading to the finding of these states are conducted in
the near future.

Predictions for such states of molecular nature were done in
Ref.~\cite{Wu:2010jy}, but lacking the present information to
establish an origin for the energies, only approximate masses could be
predicted. The success of Ref.~\cite{Xiao:2019aya} to describe the
experimental data of Ref.~\cite{Aaij:2019vzc} encourages us to use the
same formalism, adapted to the analogous states of
Ref.~\cite{Xiao:2019aya} with $c\bar{c}$ and a strange quark, to make
predictions for hidden-charm strange isoscalar molecules. For this purpose we
use the same scheme as in Ref.~\cite{Xiao:2013yca}, implementing
heavy quark spin symmetry (HQSS) in the interaction of the coupled channels
and use dimensional regularized loops with the same subtraction
constant employed in Ref.~\cite{Xiao:2019aya}.

The work of Ref. \cite{Wu:2010jy} has been complemented recently to
evaluate decay modes and rates of the states found in
Refs. \cite{Wu:2010jy,Wu:2010vk}, estimating also the uncertainties
\cite{Shen:2019evi}.

\section{Formalism}

We study states that can couple to $J/\psi \Lambda$, the channel
where, by analogy to the $J/\psi p$ of Ref.~\cite{Aaij:2019vzc}, the
new states could be observed. Thus we study states with isospin
$I=0$. The spin, however, can be 1/2 or 3/2 with negative parity for
the $S-$wave interaction that we shall consider.

The couple channels considered are:
\begin{itemize}
  
\item[i)] $J=1/2,\ I=0$

$\eta_c \Lambda$, $J/\psi \Lambda$, $\bar{D} \Xi_c$, $\bar{D}_s \Lambda_c$, 
$\bar{D} \Xi_c'$, $\bar{D}^* \Xi_c$, $\bar{D}^*_s \Lambda_c$, $\bar{D}^* \Xi_c'$, $\bar{D}^* \Xi_c^*$.

\item[ii)] $J=3/2,\ I=0$

$J/\psi \Lambda$, $\bar{D}^* \Xi_c$, $\bar{D}_s \Lambda_c$, $\bar{D}^* \Xi_c'$, $\bar{D} \Xi_c^*$, $\bar{D}^* \Xi_c^*$.

\end{itemize}  
In addition, $\bar{D}^* \Xi_c^*$ could also couple to  $J=5/2$ in $S-$wave.

\subsection{Lowest order HQSS constraints}

We take into account the lowest order (LO) constraints of
HQSS~\cite{Isgur:1989vq,Neubert:1993mb,MW00}, which states that 
interactions should be independent of the spin of the heavy quark ($Q$), up
to corrections of the order of ${\cal O}(\Lambda_{\rm QCD}/m_Q)$,
with $m_Q$ the heavy quark mass. The
spin-dependent interactions are proportional to the chromomagnetic
moment of the heavy quark, and hence, they are of the order of $1/m_Q$. The
total angular momentum $\vec{J}$ of the hadron is always a conserved
quantity, but in this case the spin of the heavy quark $\vec{S}_Q$ is also
conserved in the $m_Q\to \infty$ limit. Consequently, the spin of the
light degrees of freedom $\vec{S}_l=\vec{J}-\vec{S}_Q$ is a conserved quantity in that limit.

Here we follow the same formalism as in Ref. \cite{Xiao:2013yca} which
we briefly describe below. As we have seen, in the 
 isoscalar hidden-charm strange sector, there are 16 orthogonal states in the
physical basis composed of meson-baryon $S-$wave pairs. Next we 
introduce a different basis, that we will call HQSS basis, for which
it is straightforward to implement the LO HQSS constraints. In the
HQSS basis we will classify the states in terms of the quantum
numbers,
\begin{itemize}
\item $J=1/2,3/2$ and $5/2$: total angular momentum of the meson-baryon system

  \item ${\cal L}=1/2$ and $3/2$, and $S_{c \bar{c}}=0$ and 1: total
    angular momentum of the light-quarks and the $c \bar{c}$ subsystems, respectively.

 \item $\ell_M=0~[\eta_c,J/\psi]$ and $1/2~[D_{(s)}^{(*)}]$, and $\ell_B=1/2~[\Lambda]$, $0~[ \Xi_c]$ and $1~[
  \Xi_c^{\prime,*}]$\footnote{Throughout  this work we use
   $\Xi_c^{\prime,*}$  to refer to  $\Xi_c^\prime$ or $\Xi_c^*$.}: total angular momentum of the light quarks in the
     meson and in the baryon, respectively. Note that for the set of
     meson-baryon states considered here to construct the 
 isoscalar hidden-charm strange   sector, these quantum
     numbers also determine the total angular momentum of the heavy quarks
     in the meson and baryon $s_M$ and $s_B$, respectively.
 
\item $S=0~[D^{(*)}]$ and $-1~[D_s^{(*)}]$: strangeness of the
  meson, which also fixes that of the baryon since the total
  strangeness of the meson-baryon pair must  be $-1$.
\end{itemize}
Note that we assume that all orbital angular momenta are zero, since we are
dealing with ground state baryons, and that we have considered that the $\Xi_c$ and $\Xi_c^\prime$ baryons
are HQSS states with well defined  $\ell_B=0$ and 1,
respectively. This latter approximation seems to be quite
accurate~\cite{Bowler:1996ws,Albertus:2003sx}, and it is also implicitly  assumed in
the Review of Particle Physics~\cite{Tanabashi:2018oca}.

The approximate HQSS of QCD leads (neglecting ${\cal
  O}(\Lambda_{QCD}/m_Q)$ corrections) to:
\begin{equation}
 _{(\ell_M'\ell_B'S')}\big\langle S'_{c\bar c} {\cal L}'; J'\,|H^{QCD}|
S_{c\bar c} {\cal L}; J,\,  \big \rangle_{(\ell_M\ell_BS)} 
= \,  \delta_{JJ'}\delta_{S'_{c\bar c}S_{c\bar c} }
\delta_{{\cal L}{\cal L}'}  \big\langle \ell_M'\ell_B' S' ||H^{QCD}  || \ell_M\ell_B S \big\rangle_ {\cal L} \label{eq:hqs}
\end{equation}
Hence,  the reduced matrix elements neither depend on $S_{c\bar c}$
nor on $J$,
because QCD dynamics is invariant under separate spin rotations of the
charm quark and antiquark. Thus, one can transform a $c \bar{c}$ spin
singlet state into a spin triplet state by means of a rotation that
commutes with $H^{QCD}$, i.e. a zero cost of energy. Thus, we have a total of 11 unknown low energy terms
(LET's):
\begin{itemize}
\item One LET associated to ${\cal L}=3/2$
\begin{eqnarray}
\hat\lambda &=& \big\langle \ell_M'=1/2,\, \ell_B'=1,\,S'=0
 ||H^{QCD}  || \ell_M=1/2,\, \ell_B=1,\,S=0
 \big\rangle_{{\cal L}=3/2}
\end{eqnarray}
\item Ten LET's associated to ${\cal L}=1/2$ that form a $4\times 4$
  symmetric matrix, $\hat\mu_{ij}$, defined as
  \begin{eqnarray}
    \hat\mu_{ij} &=& \langle j ||H^{QCD}  || i \rangle_{{\cal L}=1/2}
    \nonumber \\
    |i \rangle, |j \rangle &= & | \ell_M=0,\, \ell_B=1/2,\,S=0
 \big\rangle_{{\cal L}=1/2},\, | \ell_M=1/2,\, \ell_B=0,\,S=0
 \big\rangle_{{\cal L}=1/2}, \nonumber \\
 &&  | \ell_M=1/2,\, \ell_B=0,\,S=1
 \big\rangle_{{\cal L}=1/2}\,, | \ell_M=1/2,\, \ell_B=1,\,S=0
 \big\rangle_{{\cal L}=1/2} 
\end{eqnarray}
\end{itemize}
In the HQSS basis, the  $H^{QCD}$ is a block diagonal matrix,
i.e, up to ${\cal  O}(\Lambda_{QCD}/m_Q)$ corrections, $H^{QCD}=
{\rm Diag}(\hat\mu,\hat\mu,\hat\mu,\hat\lambda,\hat\lambda,\hat\lambda,\hat\lambda)$,
where $\hat\mu$ and $\hat\lambda$ are symmetric matrices of
dimension 4 and 1, respectively. This represents an enormous
simplification, since a  16$\times$16 symmetric matrix has, in
principle, 136 independent elements.

To exploit Eq.~(\ref{eq:hqs}), one should express  the
isoscalar strange hidden-charm uncoupled meson--baryon states in terms
of the HQSS basis. The two basis are related by a Racah rotation, which is discussed in
detail in Ref.~\cite{Xiao:2013yca},
\begin{eqnarray}
  |\ell_M s_M j_M S; \ell_B s_B j_B; J\big\rangle &=& \sum_{{\cal L},S_{c\bar
      c}} \left[(2S_{c\bar
      c}+1)(2{\cal L}+1)(2j_M+1)(2j_B+1)\right]^{\frac12}\nonumber \\
  &\times& \left\{\begin{array}{ccc}\ell_M & \ell_B & {\cal L}\cr
  s_M & s_B &S_{c\bar c}\cr
  j_M & j_B & J\end{array} \right\}  | {\cal L} S_{c\bar
    c}; J \big\rangle_{(\ell_M\ell_BS)} 
\end{eqnarray}
where the angular momenta of the light and heavy degrees of freedom in
the meson $\ell_M$ and $s_M$, and in the  baryon $\ell_B$ and $s_B$, together with the
meson ($j_M$) and baryon ($j_B$) spins are coupled, by means of the
$9j-$symbol~\cite{Rose}, to ${\cal L}$,  $S_{c\bar c}$ and $J$,
respectively,  in the
  HQSS basis. At the same time, looking at the columns of the $9j$,
  $\ell_M$ and $\ell_B$, $s_M$ and $s_B$, and ${\cal L}$ and  $S_{c\bar
    c}$ are coupled to $j_M$, $j_B$ and $J$, respectively.

From Eq.~(\ref{eq:hqs}), we find that in the
isoscalar strange hidden-charm sector  the most general interactions 
  compatible with LO HQSS read 
\newpage  
  \begin{itemize}
\item $J=1/2$, $I=0$
\[
\left. \phantom{(}
\begin{array}{ccccccccc}
\phantom{ \sqrt{\frac{2}{3}} \text{$\hat\mu_{14}$}} & \phantom{-\frac{\sqrt{2} \text{$\hat\mu_{14}$}}{3}} & \phantom{-\sqrt{\frac{2}{3}} \text{$\hat\mu_{24}$}} & \phantom{-\sqrt{\frac{2}{3}} \text{$\hat\mu_{34}$}} &
   \phantom{\frac{1}{3} \sqrt{\frac{2}{3}} (\text{$\hat\mu_4$}-\hat\lambda )} & \phantom{\frac{\sqrt{2} \text{$\hat\mu
   $24}}{3}} & \phantom{\frac{\sqrt{2} \text{$\hat\mu_{34}$}}{3}} & \phantom{\frac{1}{9} \sqrt{2} (\hat\lambda -\text{$\hat\mu_4$})} & \phantom{\frac{1}{9} (\hat\lambda +8
   \text{$\hat\mu_4$})} \\
\eta_c \Lambda & J/\psi \Lambda &  \bar D \Xi_c &  \bar D_s \Lambda_c &  \bar D \Xi'_c &  \bar D^* \Xi_c
  &  \bar D^*_s  \Lambda_c &  \bar D^* \Xi'_c  &  \bar D^* \Xi^*_c 
\end{array}
\right. \phantom{)}
\]
\begin{equation}
\left( \begin{array}{ccccccccc}
\text{$\hat\mu_1$} & 0 & -\frac{\text{$\hat\mu_{12}$}}{2} & -\frac{\text{$\hat\mu_{13}$}}{2} & \frac{\text{$\hat\mu_{14}$}}{2} & \frac{\sqrt{3}
   \text{$\hat\mu_{12}$}}{2} & \frac{\sqrt{3} \text{$\hat\mu_{13}$}}{2} & \frac{\text{$\hat\mu_{14}$}}{2 \sqrt{3}} & \sqrt{\frac{2}{3}} \text{$\hat\mu_{14}$} \\ \\
 0 & \text{$\hat\mu_1$} & \frac{\sqrt{3} \text{$\hat\mu_{12}$}}{2} & \frac{\sqrt{3} \text{$\hat\mu_{13}$}}{2} & \frac{\text{$\hat\mu_{14}$}}{2 \sqrt{3}}
   & \frac{\text{$\hat\mu_{12}$}}{2} & \frac{\text{$\hat\mu_{13}$}}{2} & \frac{5 \text{$\hat\mu_{14}$}}{6} & -\frac{\sqrt{2} \text{$\hat\mu_{14}$}}{3} \\ \\
 -\frac{\text{$\hat\mu_{12}$}}{2} & \frac{\sqrt{3} \text{$\hat\mu_{12}$}}{2} & \text{$\hat\mu_2$} & \text{$\hat\mu_{23}$} & 0 & 0 & 0 & \frac{\text{$\hat\mu_{24}$}}{\sqrt{3}} & -\sqrt{\frac{2}{3}} \text{$\hat\mu_{24}$} \\ \\
 -\frac{\text{$\hat\mu_{13}$}}{2} & \frac{\sqrt{3} \text{$\hat\mu_{13}$}}{2} & \text{$\hat\mu_{23}$} & \text{$\hat\mu_3$} & 0 & 0 & 0 & \frac{\text{$\hat\mu_{34}$}}{\sqrt{3}} & -\sqrt{\frac{2}{3}} \text{$\hat\mu_{34}$} \\ \\
 \frac{\text{$\hat\mu_{14}$}}{2} & \frac{\text{$\hat\mu_{14}$}}{2 \sqrt{3}} & 0 & 0 & \frac{1}{3} (2 \hat\lambda +\text{$\hat\mu_4$}) &
   \frac{\text{$\hat\mu_{24}$}}{\sqrt{3}} & \frac{\text{$\hat\mu_{34}$}}{\sqrt{3}} & -\frac{2 (\hat\lambda -\text{$\hat\mu_4$})}{3 \sqrt{3}} &
   \frac{1}{3} \sqrt{\frac{2}{3}} (\text{$\hat\mu_4$}-\hat\lambda ) \\ \\
 \frac{\sqrt{3} \text{$\hat\mu_{12}$}}{2} & \frac{\text{$\hat\mu_{12}$}}{2} & 0 & 0 & \frac{\text{$\hat\mu_{24}$}}{\sqrt{3}} & \text{$\hat\mu_2$} &
   \text{$\hat\mu_{23}$} & \frac{2 \text{$\hat\mu_{24}$}}{3} & \frac{\sqrt{2} \text{$\hat\mu_{24}$}}{3} \\ \\
 \frac{\sqrt{3} \text{$\hat\mu_{13}$}}{2} & \frac{\text{$\hat\mu_{13}$}}{2} & 0 & 0 & \frac{\text{$\hat\mu_{34}$}}{\sqrt{3}} & \text{$\hat\mu_{23}$} &
   \text{$\hat\mu_3$} & \frac{2 \text{$\hat\mu_{34}$}}{3} & \frac{\sqrt{2} \text{$\hat\mu_{34}$}}{3} \\ \\
 \frac{\text{$\hat\mu_{14}$}}{2 \sqrt{3}} & \frac{5 \text{$\hat\mu_{14}$}}{6} & \frac{\text{$\hat\mu_{24}$}}{\sqrt{3}} & \frac{\text{$\hat\mu_{34}$}}{\sqrt{3}} & -\frac{2 (\hat\lambda -\text{$\hat\mu_4$})}{3 \sqrt{3}} & \frac{2 \text{$\hat\mu_{24}$}}{3} & \frac{2 \text{$\hat\mu_{34}$}}{3} &
   \frac{1}{9} (2 \hat\lambda +7 \text{$\hat\mu_4$}) & \frac{1}{9} \sqrt{2} (\hat\lambda -\text{$\hat\mu_4$}) \\ \\
 \sqrt{\frac{2}{3}} \text{$\hat\mu_{14}$} & -\frac{\sqrt{2} \text{$\hat\mu_{14}$}}{3} & -\sqrt{\frac{2}{3}} \text{$\hat\mu_{24}$} &
   -\sqrt{\frac{2}{3}} \text{$\hat\mu_{34}$} & \frac{1}{3} \sqrt{\frac{2}{3}} (\text{$\hat\mu_4$}-\hat\lambda ) & \frac{\sqrt{2} \text{$\hat\mu_{24}$}}{3} & \frac{\sqrt{2} \text{$\hat\mu_{34}$}}{3} & \frac{1}{9} \sqrt{2} (\hat\lambda -\text{$\hat\mu_4$}) & \frac{1}{9} (\hat\lambda +8
   \text{$\hat\mu_4$}) \\
\end{array} \right)
\label{eq:ji11}
\end{equation}

\item  $J=3/2$, $I=0$
\[
\left. \phantom{(}
\begin{array}{cccccc}
\phantom{\frac{\sqrt{5} \text{$\hat\mu_{14}$}}{3} } & \phantom{\frac{\sqrt{5} \text{$\hat\mu_{24}$}}{3}} & \phantom{\frac{\sqrt{5} \text{$\hat\mu_{34}$}}{3}} & \phantom{\frac{1}{9}
   \sqrt{5} (\hat\lambda -\text{$\hat\mu_4$})} & \phantom{\frac{1}{3} \sqrt{\frac{5}{3}} (\text{$\hat\mu_4$}-\hat\lambda )} & \phantom{\frac{1}{9} (4 \hat\lambda +5
   \text{$\hat\mu_4$})}  \\
 J/\psi \Lambda &  \bar D^* \Xi_c &  \bar D_s^* \Lambda_c 
&  \bar D^* \Xi'_c  &  \bar D \Xi^*_c  & \bar D^* \Xi_c^*
\end{array}
\right. \phantom{)}
\]
\begin{equation}
\left(
\begin{array}{cccccc}
 \text{$\hat\mu_1$} & \text{$\hat\mu_{12}$} & \text{$\hat\mu_{13}$} & -\frac{\text{$\hat\mu_{14}$}}{3} & \frac{\text{$\hat\mu_{14}$}}{\sqrt{3}} &
   \frac{\sqrt{5} \text{$\hat\mu_{14}$}}{3} \\ \\
 \text{$\hat\mu_{12}$} & \text{$\hat\mu_2$} & \text{$\hat\mu_{23}$} & -\frac{\text{$\hat\mu_{24}$}}{3} & \frac{\text{$\hat\mu_{24}$}}{\sqrt{3}} &
   \frac{\sqrt{5} \text{$\hat\mu_{24}$}}{3} \\ \\
 \text{$\hat\mu_{13}$} & \text{$\hat\mu_{23}$} & \text{$\hat\mu_3$} & -\frac{\text{$\hat\mu_{34}$}}{3} & \frac{\text{$\hat\mu_{34}$}}{\sqrt{3}} &
   \frac{\sqrt{5} \text{$\hat\mu_{34}$}}{3} \\ \\
 -\frac{\text{$\hat\mu_{14}$}}{3} & -\frac{\text{$\hat\mu_{24}$}}{3} & -\frac{\text{$\hat\mu_{34}$}}{3} & \frac{1}{9} (8 \hat\lambda +\text{$\hat\mu_4$}) &
   \frac{\hat\lambda -\text{$\hat\mu_4$}}{3 \sqrt{3}} & \frac{1}{9} \sqrt{5} (\hat\lambda -\text{$\hat\mu_4$}) \\ \\
 \frac{\text{$\hat\mu_{14}$}}{\sqrt{3}} & \frac{\text{$\hat\mu_{24}$}}{\sqrt{3}} & \frac{\text{$\hat\mu_{34}$}}{\sqrt{3}} & \frac{\hat\lambda
   -\text{$\hat\mu_4$}}{3 \sqrt{3}} & \frac{1}{3} (2 \hat\lambda +\text{$\hat\mu_4$}) & \frac{1}{3} \sqrt{\frac{5}{3}} (\text{$\hat\mu_4$}-\hat\lambda
   ) \\ \\
 \frac{\sqrt{5} \text{$\hat\mu_{14}$}}{3} & \frac{\sqrt{5} \text{$\hat\mu_{24}$}}{3} & \frac{\sqrt{5} \text{$\hat\mu_{34}$}}{3} & \frac{1}{9}
   \sqrt{5} (\hat\lambda -\text{$\hat\mu_4$}) & \frac{1}{3} \sqrt{\frac{5}{3}} (\text{$\hat\mu_4$}-\hat\lambda ) & \frac{1}{9} (4 \hat\lambda +5
   \text{$\hat\mu_4$}) \\
\end{array}
\right)
\label{eq:ji31}
\end{equation}

\item $J=5/2$, $I=0$
  \begin{equation}
    {\bar D}^* \Xi_c^*: \, \hat\lambda
    \label{eq:ji51}
\end{equation}
\end{itemize}

\subsection{Coupled-channels unitarity and the Bethe-Salpeter equation (BSE)}  
  
For each $J$, we use the BSE in coupled channels
\begin{equation}
T = [1 - V \, G]^{-1}\, V, \label{eq:BS}
\end{equation}
where $V_{ij}$ is the two-particle irreducible amplitude (potential),
and $G$ is a diagonal matrix  constructed out of the loop functions
for intermediate meson baryon states.

For any choice of $V$ and of the renormalization scheme adopted to
evaluate ${\rm Re}[G]$ on the real axis, we find that above the lowest
of the meson-baryon thresholds, the discontinuity of $T^{-1}$ is equal
to that of $-G$, fulfilling in this way exact unitary in
coupled channels~\cite{Nieves:1999bx}. Here we implement LO HQSS
relations in the kernel $V$, which is taken from Eqs.~\eqref{eq:ji11}
and \eqref{eq:ji31} for $J=1/2$ and $J=3/2$ respectively, while for
$\bar{D}^* \Xi_c^* \to \bar{D}^* \Xi_c^*$, $J=5/2$, the single-channel
potential is $\hat\lambda$ (Eq. \eqref{eq:ji51}). On the other hand, we
calculate the loop functions in dimensional regularization using the
formula of Refs.~\cite{Oller:2000fj,Oset:2001cn}, and take advantage
of our previous work of Ref.~\cite{Xiao:2019aya} to fix $a(\mu=1\,{\rm
  GeV})=-2.09$. This subtraction constant was determined in this
latter reference to agree with the average experimental mass of the
three non-strange pentaquarks reported by the LHCb Collaboration in
\cite{Aaij:2019vzc}.

\subsection{Interactions from the local hidden gauge (LHG) approach}

\label{susec:theory}

HQSS gives us the structure of $V$ in terms of
the irreducible matrix elements of the interaction, but the strength
is not given. Then, as in Ref.~\cite{Xiao:2013yca}, we take these
matrix elements using an extension of the LHG
approach~\cite{Bando:1984ej,Bando:1987br,Meissner:1987ge,Nagahiro:2008cv}.
This picture is based on the exchange of vector mesons between the
meson and the baryon. Actually, it has a direct connection
with HQSS. Indeed, the dominant terms are those
which exchange light vector mesons ($\rho,\ \omega,\ \phi,\ K^*$),
since the exchange of heavy mesons is suppressed by large masses in the propagators
involving them. The $c$ and $\bar c$ quarks in this case act as spectators in the
interaction, which then does not depend upon them, and automatically
the independence on the spin of the heavy quarks (or any other
property) is fulfilled. This is not the case when heavy vectors are
exchanged, and neither should it be, since these terms are of order
($1/m_Q^2$) and then sub-leading in the $1/m_Q$ counting.

The evaluation of the matrix elements in Ref.~\cite{Xiao:2013yca} was
done following the work of \cite{Wu:2010jy} where an extrapolation to
SU(4) was done. In between it has become apparent that the use of
SU(4) is unnecessary and that a perfect counting stems from the
overlap of quarks of suitable wave functions that were used in
Ref.~\cite{Debastiani:2017ewu}. We use the same procedure here and
explain it below adapted to the present case.

We first write down the baryon wave functions that we use, where we
single out the heavy quarks, and impose the spin-flavour symmetry in
the light quarks:

\begin{itemize}

\item[1)] $\Lambda$: $\frac{1}{\sqrt{2}} (\phi_{MS}\chi_{MS} +
  \phi_{MA}\chi_{MA})$, where $\chi_{MS},\ \chi_{MA}$ are the mixed
  symmetric and mixed antisymmetric representations of the spin 1/2 of
  the baryons~\cite{close}. $\phi_{MS},\ \phi_{MA}$ are the flavour
  mixed symmetric and antisymmetric wave functions for the SU(3)
  baryons. Here, we must divert from the prescription of \cite{close}
  to make the phase convention consistent with the use of chiral
  Lagrangians and we follow the convention of
  Refs. \cite{Miyahara:2016yyh,Pavao:2017cpt}. Hence, with the
  symmetry in the last two particles, we have
\begin{align}
\phi_{MS}&=-\frac{1}{\sqrt{2}} \Big[ \frac{uds-dus}{\sqrt{2}} + \frac{usd-dsu}{\sqrt{2}} \Big], \\
\phi_{MA}&=-\frac{1}{\sqrt{6}} \Big[ \frac{usd-dsu}{\sqrt{2}} + \frac{dus-uds}{\sqrt{2}} -2 \frac{sdu-sud}{\sqrt{2}} \Big].
\end{align}

\item[2)] $\Lambda_c^+: c\,\frac{1}{\sqrt{2}} (ud-du) \chi_{MA}$.

\item[3)] $\Xi_c^+: c\,\frac{1}{\sqrt{2}} (us-su) \chi_{MA}$ and
$\Xi_c^0: c\,\frac{1}{\sqrt{2}} (ds-sd) \chi_{MA}$.

\item[4)] $\Xi_c^{'+}: c\,\frac{1}{\sqrt{2}} (us+su) \chi_{MS}$ and 
$\Xi_c^{'0}:  c\,\frac{1}{\sqrt{2}} (ds+sd) \chi_{MS}$.

\item[5)] $\Xi_c^{*+}: c\,\frac{1}{\sqrt{2}} (us+su) \chi_{S}$ and 
  $\Xi_c^{*0}: c\,\frac{1}{\sqrt{2}} (ds+sd) \chi_{S}$.
\end{itemize}
with  $\chi_S$ the 3/2 spin symmetric wave function \cite{close}. The
light-quark parts of the above wave-functions are consistent with
$\ell_B=0$ for $\Lambda_c$ and $\Xi_c$, and $\ell_B=1$ for $\Xi_c^\prime$
and $\Xi_c^*$. On the other hand, our conventions for the  isospin
doublets are $(D^+,\ -D^0)$, $(\bar{D}^0,\ D^-)$,
$(\Xi_c^+,\ \Xi_c^0)$. Thus, for instance,
we have that the isoscalar $\bar{D} \Xi_c$ wave  function is
\begin{equation}
|\bar{D} \Xi_c,\ I=0\rangle = \frac{1}{\sqrt{2}} |\bar{D}^0 \Xi_c^0 - D^- \Xi_c^+ \rangle.
\end{equation}
\begin{figure}[H]
\centering
\includegraphics[scale=0.55]{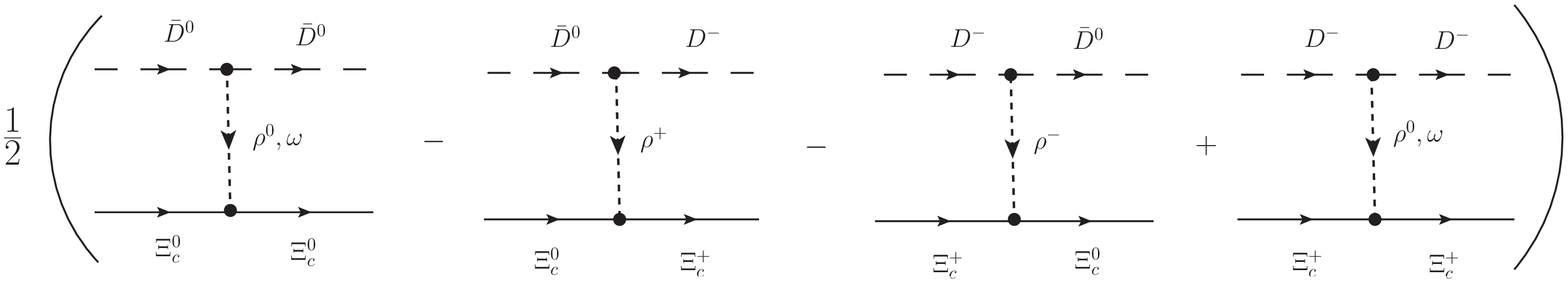}
\caption{Diagrams for $\bar{D}\Xi_c\to\bar{D}\Xi_c$ transitions.}
\label{fig:tran}
\end{figure}
We must then evaluate the matrix elements of Fig.~\ref{fig:tran}.  Let
us evaluate explicitly the first one of these diagrams. The
lower vertex is easily evaluated using the Lagrangian
\begin{equation}
{\cal L}_{VBB} \equiv g \left\{ 
\begin{aligned}
& \frac{1}{\sqrt{2}} (u\bar{u} - d\bar{d}),\ \rho^0 \\
& \frac{1}{\sqrt{2}} (u\bar{u} + d\bar{d}),\ \omega
\end{aligned} \right\},
\end{equation}
which implicitly assumes that the $\gamma^\mu$ vertex for low energy baryons is converted into $\gamma^0$ which is unity in this case. Hence, we have
\begin{equation}
\frac{1}{\sqrt{2}} \Big\langle c(ds-sd) \chi_{MA} \Big| g \left\{
\begin{array}{c}
\frac{1}{\sqrt{2}} (u\bar{u} - d\bar{d}) \cr\cr
\frac{1}{\sqrt{2}} (u\bar{u} + d\bar{d})
\end{array} \right\} \Big| c(ds-sd) \chi_{MA} \Big\rangle = \frac{g}{\sqrt{2}} \left\{
\begin{array}{cc}
 -1, & \rho^0 \cr\cr
\phantom{-} 1, &\omega
\end{array} \right\},
\end{equation}
where $g$ is the coupling of the LHG, $g=M_V /2 f_\pi$ ($M_V \approx
800\mev,\ f_\pi = 93 \mev$). The upper vertex can be evaluated in a
similar way using wave functions for the mesons as shown in
Ref.~\cite{Sakai:2017avl} (see section IIA of that reference), but it
is also shown there that for practical reasons it is easier to get the
vertex from the Lagrangian
\begin{equation}
{\cal L}_{VPP}= -ig ~\langle [P,\partial_{\mu}P]V^{\mu}\rangle, \label{eq:lvpp}
\end{equation}
with $\langle...\rangle$ the matrix trace, and 
\begin{equation}
P = \left(
\begin{array}{cccc}
\frac{\pi^0}{\sqrt{2}}+\frac{\eta}{\sqrt{3}}+\frac{\eta'}{\sqrt{6}}        &\pi^+     & K^{+}    &\bar{D}^{0}  \\
\pi^-      & -\frac{\pi^0}{\sqrt{2}}+\frac{\eta}{\sqrt{3}}+\frac{\eta'}{\sqrt{6}}      & K^{0}         &D^{-}  \\
K^{-}      & \bar{K}^{0}       & -\frac{\eta}{\sqrt{3}}+\sqrt{\frac{2}{3}} \eta'        & D^{-}_s\\
D^{0}   &  D^{+}     &  D^{+}_s    &  \eta_c \\
\end{array}
\right) \ ,
\end{equation}
\begin{equation}
V_\mu=\left(
\begin{array}{cccc}
\frac{\rho^0}{\sqrt{2}}+\frac{\omega}{\sqrt{2}} & \rho^+ & \quad K^{*+}\quad & \quad \bar D^{*0} \\ & & & \\
 \rho^{-} & -\frac{\rho^0}{\sqrt{2}} + \frac{\omega}{\sqrt{2}} & K^{*0} & D^{*-} \\
 & & & \\
  K^{*-} & \bar K^{*0} & \phi & D_s^{*-} \\
  & & & \\
D^{*0} & D^{*+} & D_s^{*+} & J/\psi \\ 
\end{array}
\right)_\mu \ .
\end{equation}
One finds readily that the vertex for $\bar{D}^0 \bar{D}^0 \rho\ (\omega)$ is given by
\begin{equation}
t=-g (p^\mu + p'^\mu) \epsilon_\mu \left\{
\begin{aligned}
& \frac{1}{\sqrt{2}},\ \rho^0 \\
& \frac{1}{\sqrt{2}},\ \omega
\end{aligned} \right\}.
\end{equation}
with $\epsilon_\mu$ the polarization vector of the exchanged virtual
vector meson. As mentioned above, the baryon vertex selects the
exchange of the temporal component. We thus see that in the first and
last diagrams of Fig.~\ref{fig:tran} there is a cancellation of the
$\rho$ and $\omega$ contributions, assuming equal $\rho$ and $\omega$
masses, and the contribution only comes from the two crossed terms,
second and third diagrams of Fig.~\ref{fig:tran}. Repeating the
procedure for these terms we finally find
\begin{equation}
\hat\mu_2=- g^2 (p^0+p^{\prime\, 0}) \frac{1}{m_V^2} = -\frac{1}{4 f_\pi^2} (p^0+p^{\prime\, 0}),
\end{equation}
where $p^0$ and $p^{\prime\, 0}$ are the energies of the external mesons.

It is easy to see that in $\bar{D}_s \Lambda_c \to \bar{D}_s
\Lambda_c$ we cannot exchange $\rho^0,\ \omega$ since there are no
$u,\ d$ quarks in $\bar{D}_s$ and we cannot exchange $\phi$ since
there are no strange quarks in $\Lambda_c$. Thus,
\begin{equation}
\hat\mu_3=0.
\end{equation}

We can follow the same steps to find $\hat\mu_{23}$ and the other
coefficients. It will be also needed to consider $D^*_{(s)}$ vector-mesons as
external legs.  For the three vector vertex, we use
\begin{equation}
{\cal L}^{(3V)}_{III}=ig\langle [V^\nu,\partial_{\mu}V^\nu] V^\mu \rangle , \label{eq:lvvv}
\end{equation}
and it is shown in Ref.~\cite{Oset:2009vf} that in the limit of small
external three momenta, which we assume here, it has the same
structure as Eq.~\eqref{eq:lvpp} for the equivalent pseudoscalar
mesons, with $V^\mu$ playing the role of the exchanged vector, and the
additional $\vec{\epsilon}\,\vec{\epsilon}\,'$ factor, with
$\vec{\epsilon}$ and $\vec{\epsilon}\,'$ the polarization vectors of the
external vector-mesons. Indeed, it is easy to see that $V^\mu$ cannot be an
external meson. If this were the case, $\mu$ should be a space
component since $\epsilon^0=0$ for vectors at rest, then $\partial_i
V_i \ (i=1,\ 2,\ 3)$ is proportional to the three momenta, which are
neglected in this approach.

Finally we obtain
\begin{eqnarray}
\hat\mu_1 &=& \hat\mu_3 =  \hat\mu_{24} = \hat\mu_{34} = 0\\ 
\hat\mu_2 &=& \hat\mu_{23}/\sqrt{2}= \hat\mu_4=\hat\lambda=-F, \qquad
F=\frac{1}{4f^2} (p^0 + p^{\prime\,}) \\
\hat\mu_{12} &=& -\hat\mu_{13}/\sqrt{2}=\hat\mu_{14}/\sqrt{3} =
-\sqrt{\frac{2}{3}}\ \frac{m_V^2}{m_{D^*}^2}\ F,
\end{eqnarray}
where we have explicitly implemented the reduction factor
$m_V^2/m_{D^*}^2$ in the matrix elements that involve change of $c$
quark content in the mesons, which proceed in our approach via the
exchange of $D^*$. The null $\hat\mu_{34}$ and $\hat\mu_{24}$ values stem from
our neglect of pion exchange, as done in
Refs.~\cite{Xiao:2013yca,Xiao:2019aya}.

We should make here two important remarks:
\begin{itemize}
\item Up to some minus signs in off diagonal matrix elements, which
  can be re-absorbed by conveniently redefining the overall phase of the
  involved mesons and baryons, the $\bar D^{(*)}\Xi^\prime_c- \bar
  D^{(*)}\Xi^*_c$ sub-matrices here are identical to those used in
  Ref.~\cite{Xiao:2019aya} for the $\bar D^{(*)}\Sigma_c-\bar
  D^{(*)}\Sigma_c^* $ in the non-strange hidden charm sector with $I=1/2$. This is
  is the case for all $J=1/2, 3/2$ and 5/2 angular momenta and when 
  the LET's $\hat\lambda$ and $\hat\mu_4$ that appear here are
  identified to 
  $\lambda_2$ and $\mu_3$ introduced in
  Ref.~\cite{Xiao:2019aya}. Moreover all these terms are predicted to
  be equal to $-F$ within the LHG approach, as deduced from
  SU(3)-light flavor
  symmetry. The correspondence between $(\bar D^{(*)}\Xi^\prime_c)_{I=0}$ and
  $(\bar D^{(*)}\Sigma_c)_{I=3/2}$,
  and between $(\bar D^{(*)} \Xi^*_c)_{I=0}$ and $(\bar D^{(*)}\Sigma^*_c)_{I=1/2}$
  is natural because in all these
  baryons the light quarks are coupled to $\ell_B=1$, and in both cases
  the isospin meson-baryon coupling is symmetric.

  A straightforward consequence is that the pattern of seven states
  obtained in Ref.~\cite{Xiao:2019aya}, located few MeV below the
  $\bar D^{(*)}\Sigma_c^{(*)}$ thresholds, is expected to be also
  found here. We will see that seven quasi-bound $\bar
  D^{(*)}\Xi^{\prime,*}_c$ isoscalar states, SU(3) partners of the
  former ones, are dynamically generated. Two $(J=1/2,3/2)$ HQSS
  doublets and a $(J=1/2,3/2,5/2)$ HQSS triplet. The latter multiplet is
  originated from the $\bar D^*\Xi_c^*$ interaction, the heaviest
  doublet from the $\bar D^* \Xi_c^\prime$, while the other one
  corresponds to $\bar D \Xi^\prime_c$ ($J=1/2$) and $\bar D \Xi_c^*$
  ($J=3/2$). In the
  non-strange sector, the first doublet was identified in
  \cite{Xiao:2019aya} with the $P_c(4440)$ and the $P_c(4457)$ LHCb
  states, while the quasi-bound $\bar D\Sigma_c$ state had a mass and
  width compatible with those of the $P_c(4312)$.
  
  \item In the strict heavy quark limit, $\hat\mu_{12}$ and
    $\hat\mu_{13}$ also vanish within the HLG approach. Then, the
    $\bar D^{(*)} \Xi_c$ and $\bar D_s^{(*)}\Lambda_c$ decouple from
    the rest of channels in the $J=1/2$ and $J=3/2$ sectors. In both
    cases, the isoscalar interaction is determined from $\hat\mu_2$ and
    $\hat\mu_{23}$, and it reads
    \begin{equation}
      V_{\bar D^{(*)} \Xi_c,\bar D_s^{(*)}\Lambda_c } = F \left
      (\begin{array}{cc} -1 & -\sqrt{2} \cr
      -\sqrt{2} & 0\end{array} \right)
\end{equation}
\end{itemize}
Diagonalizing the above matrix, we find eigenvalues $F$ and
$-2F$. The latter one is  attractive and the corresponding
eigenvector is dominated by the $\bar D^{(*)} \Xi_c$ component,
i.e. $(\sqrt{2}\,\bar D^{(*)} \Xi_c, \bar
D_s^{(*)}\Lambda_c)/\sqrt{3}$. Hence, we
should expect the existence of a $J=1/2$ isoscalar
$\bar D
\Xi_c$ quasi-bound  state around 4337 MeV and an isoscalar $(J=1/2,3/2)$ HQSS
doublet located near the $\bar D^*\Xi_c-$threshold (4479 MeV).

On the other hand, in this $m_Q\to \infty$ limit, the $\bar
D^{(*)}\Xi^{\prime,*}_c$ interactions become diagonal in the
meson-baryon basis, with a common strength of $-F$ since
$\hat\mu_4=\hat\lambda$. Therefore, the $\bar D^{(*)}\Xi_c$ states
should be more bound than those generated from the $\bar
D^{(*)}\Xi^{\prime,*}_c$ interaction. Moreover in the latter case, all
the binding energies should be similar, as it occurred in the
non-strange case discussed in  \cite{Xiao:2019aya}. There, all states
were found around 10 MeV below the corresponding $\bar
D^{(*)}\Sigma_c^{(*)}-$thresholds.

Finally, let us note that in the non-strange hidden charm sector the
channel equivalent to $\bar D^{(*)}\Xi_c$ would be the $\bar
D^{(*)}\Lambda_c$, with $\ell_B=0$ in both cases, but with a totally
different meson-baryon isospin structure. Indeed, because of the
isospin couplings, the $\bar D^{(*)}\Lambda_c$ interaction turns out
to be repulsive ($+F$)~\cite{Xiao:2013yca}, instead of attractive as
we found here. As a consequence, we did not
find $\bar D^{(*)}\Lambda_c$ hadron molecules in our previous study of
Ref.~\cite{Xiao:2019aya} in the non-strange sector.

\section{Results}

\begin{figure}
\centering
\includegraphics[scale=0.6]{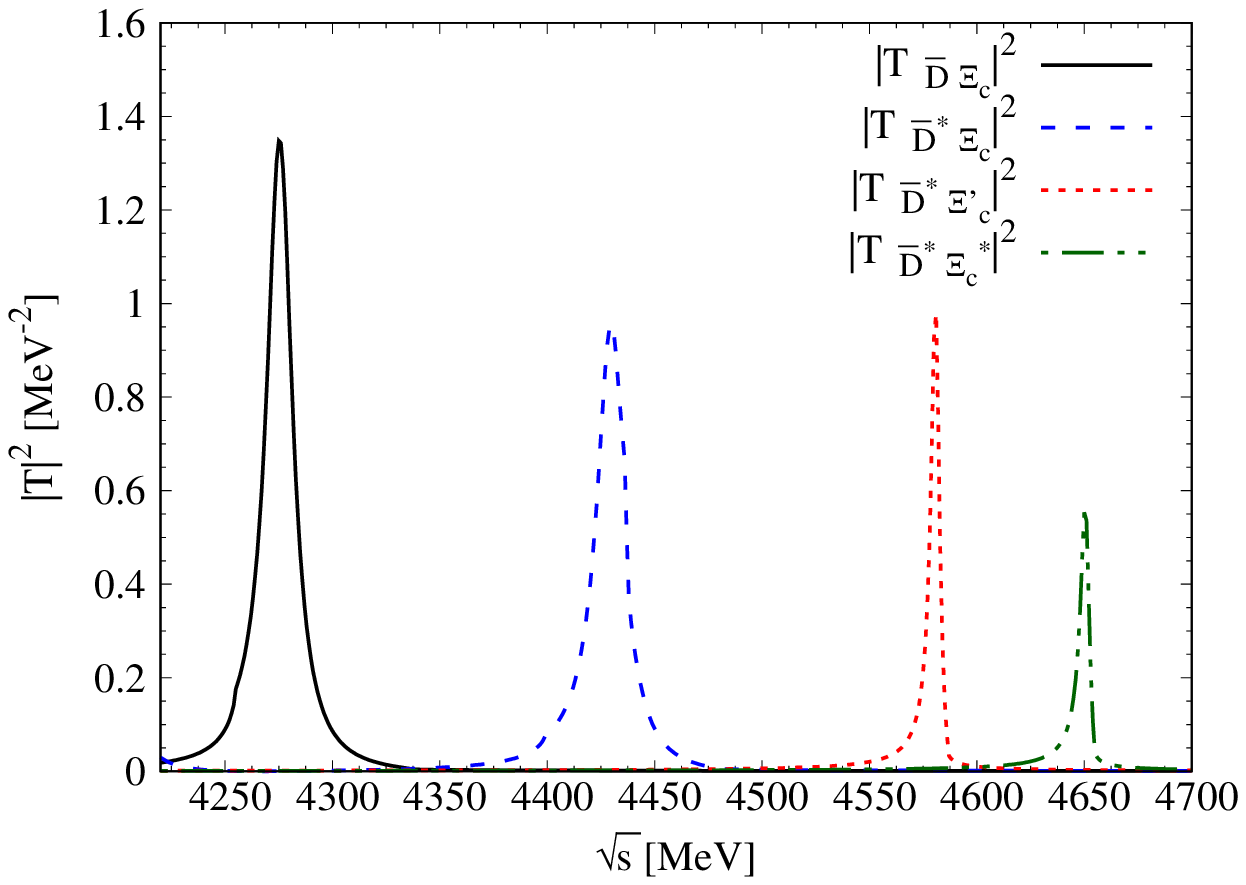}
\includegraphics[scale=0.6]{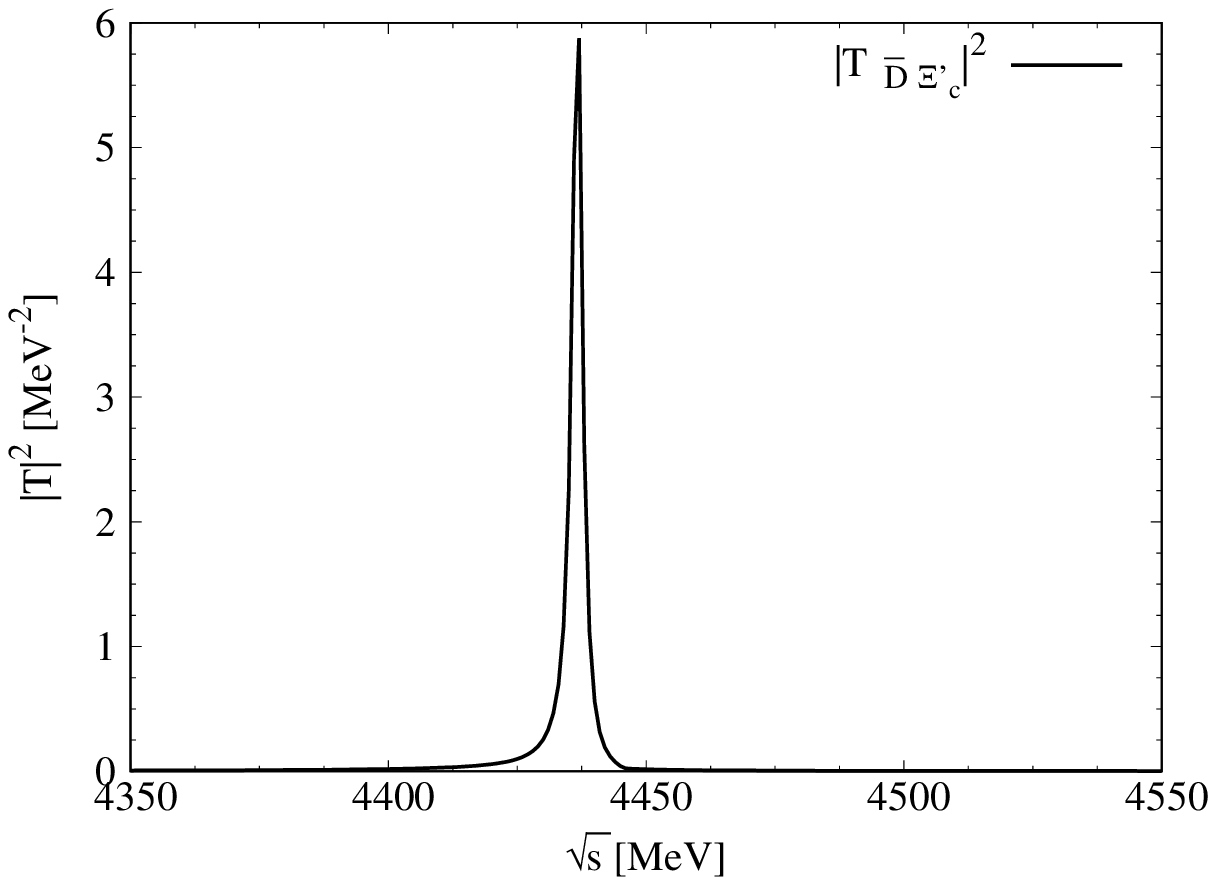}
\caption{Results of the modulus squared of some diagonal elements of
  the  amplitude matrix, as a function of $\sqrt{s}$, for the
  $J=1/2,~I=0$ sector.}
\label{fig:tsq12}
\end{figure}

In Fig.~\ref{fig:tsq12} we show the results for $|T|^2$ for the
diagonal matrix elements of Eq.~\eqref{eq:BS} and $J=1/2$. The peaks
indicate where states appear and one can also see qualitatively the
width of these states. More details can be seen by looking explicitly
for poles in the second Riemann sheet~\cite{Xiao:2013yca}, and evaluating the couplings
obtained from the amplitude close to the pole, 
\begin{equation}
T_{ij} \simeq \frac{g_i g_j}{\sqrt{s}-\sqrt{s_R}}, \qquad
\sqrt{s_R}=M + i\,\Gamma/2
\end{equation}
We show this information in Table~\ref{tab:cou12}.  We find five
states with $J=1/2$, at $(M +i\,\Gamma/2)=$ $(4276.59+i7.67)\mev$,
$(4429.89+i7.92)\mev$, $(4436.70+i1.17)\mev$,  $(4580.96+i2.44)\mev$,
and $(4650.86+i2.59)\mev$. We observe that the widths are small in all
cases. In the table we stress with thick lettering the biggest
couplings, which indicate the dominance of some channels. We see that
the dominant channels for the states reported before are
$\bar{D}\Xi_c$, $\bar{D}^*\Xi_c$, $\bar{D}\Xi_c'$, $\bar{D}^*\Xi_c'$
and $\bar{D}^*\Xi_c^*$, respectively. The two lightest states have
also some significant couplings  to the open $\bar{D}_s\Lambda_c$ and 
$\bar{D}_s^*\Lambda_c$ channels in each case, as expected from the
discussion at the end of Subsect.~\ref{susec:theory}, and that gives
rise to widths of around 15 MeV. 
The couplings of all states to
$J/\psi \Lambda$, the channel where these states are most likely to be
observed, are relatively small, but sufficiently large to provide
production rates in observable ranges, if one compares their strengths
with the ones obtained for $J/\psi N$ in the hidden charm sector
without strangeness \cite{Xiao:2013yca} where the new pentaquark peaks
have been observed \cite{Aaij:2019vzc}.

\begin{figure}
\centering
\includegraphics[scale=0.6]{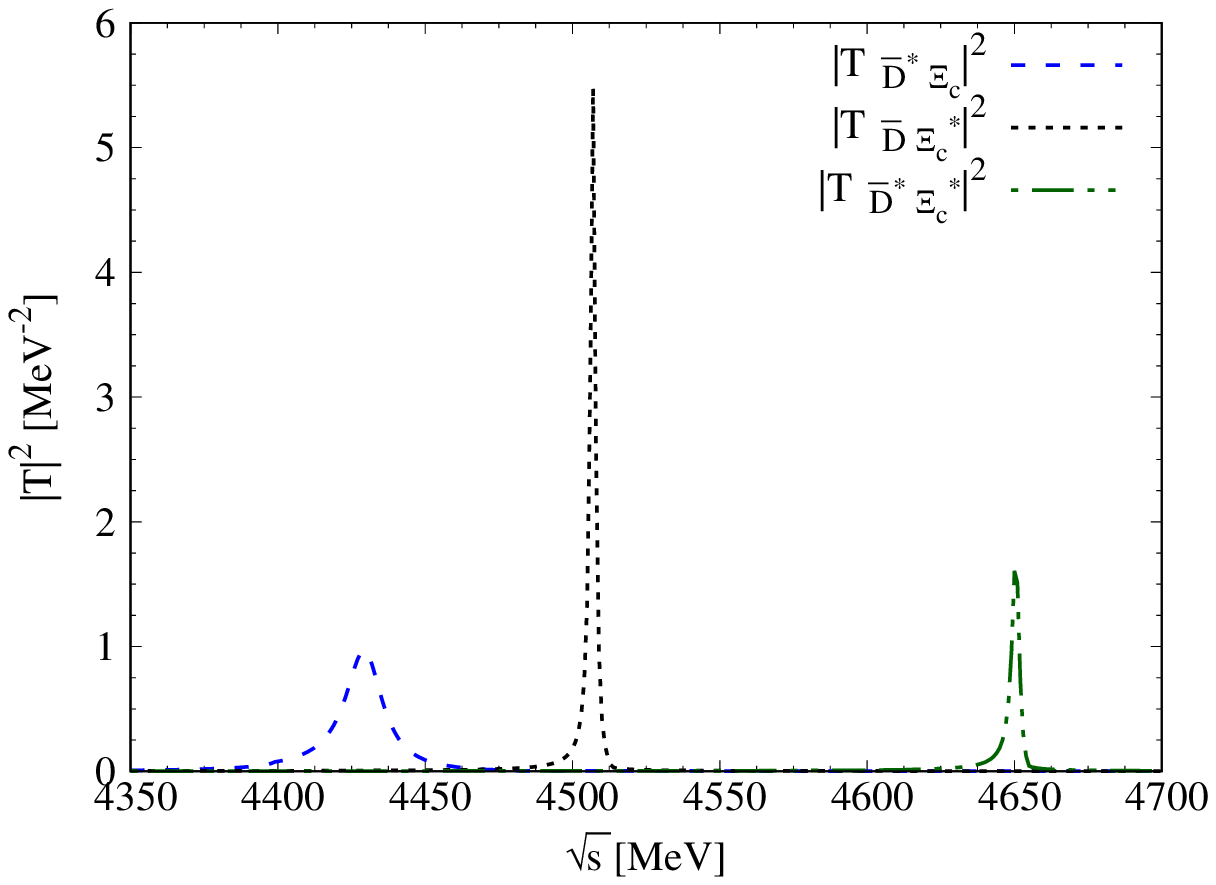}
\includegraphics[scale=0.6]{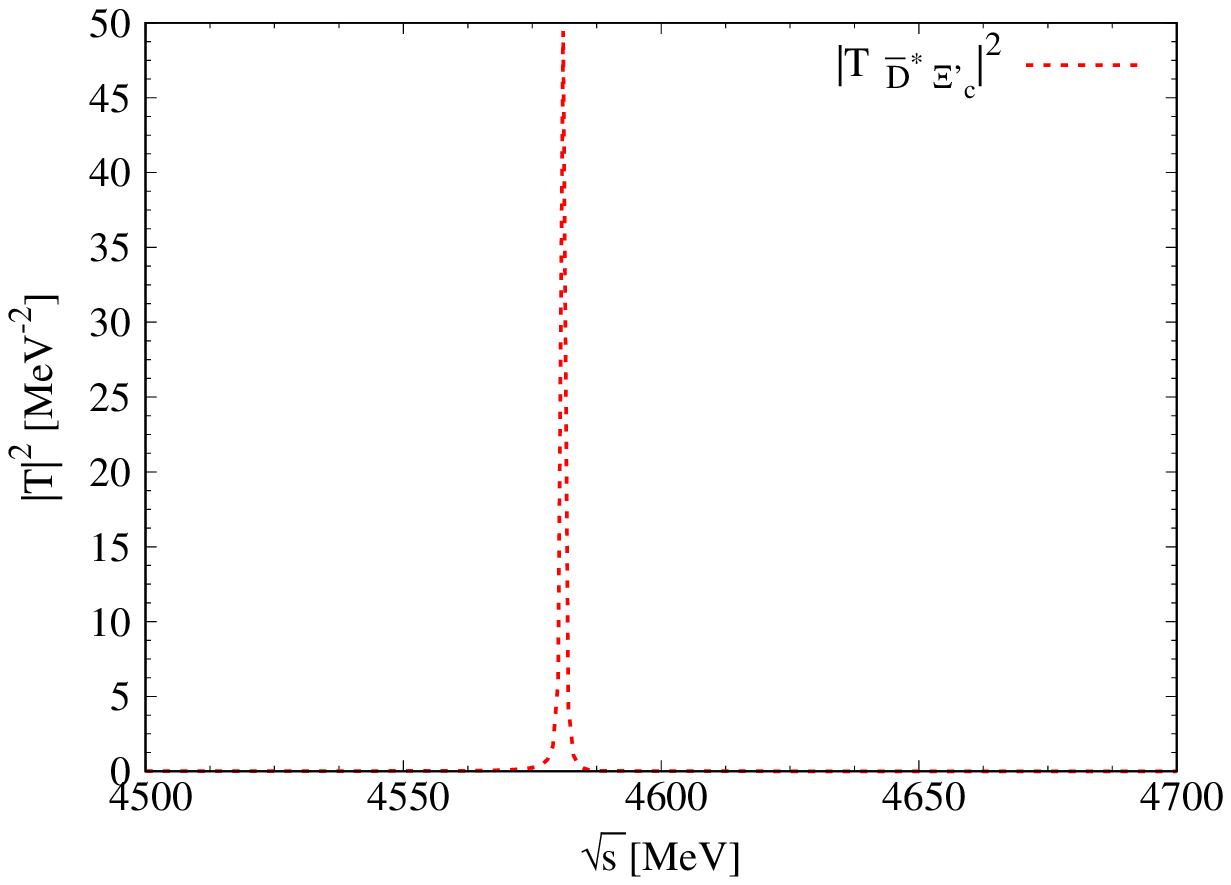}
\caption{Same as Fig.~\ref{fig:tsq12} for $J= 3/2$.}
\label{fig:tsq32}
\end{figure}

In the $J=3/2$ sector we also find states, first depicted by means of
$|T|^2$ in Fig.~\ref{fig:tsq32}. Once again, we summarize the
information of the poles and couplings in Table~\ref{tab:cou32}. Here
we find four states at $(4429.52+i7.67)\mev$,
$(4506.99+i1.03)\mev$, $(4580.96+i0.34)\mev$ and $(4650.58+i1.48)\mev$. They are again
narrow and couple mostly to the $\bar{D}^*\Xi_c$, $\bar{D}\Xi_c^*$,
$\bar{D}^*\Xi_c'$ and $\bar{D}^*\Xi_c^*$ channels, respectively. The
lightest state also couples to the open $\bar{D}^*_s\Lambda_c$
channel, as expected, which leads two a sizable width of around 15 MeV. The
couplings to $J/\psi \Lambda$ are again sufficiently large compared to
those of $J/\psi N$ in the hidden charm non-strange case, such that
the observation of peaks in the $J/\psi \Lambda$ channel should not be
a problem. We should note that the channels $\bar{D} ^*\Xi_c$,
$\bar{D}^*\Xi_c'$, $\bar{D}^*\Xi_c^*$ participate both in $J=1/2$ and
$J=3/2$. The $J=1/2$ and $J=3/2$ states obtained which couple mostly to each of these
channels are practically degenerate but have a different width. This
is very similar to the situation found in Ref.~\cite{Xiao:2019aya}
for the $P_c(4440)$ and $P_c(4457)$ states of
Ref.~\cite{Aaij:2019vzc}, which in our approach appear near degenerate
in $1/2^-,\ 3/2^-$ but with a larger width for the $1/2^-$
state. Here, the situation is similar and, although all the states are
quite narrow, the $1/2^-$ states have slightly larger widths than their
corresponding $3/2^-$ ones.

In the $J=5/2^-$ sector we have just the $\bar{D}^* \Xi_c^*$ state
with a mass $4650.56\mev$ with a zero width, which is also degenerate
with the $1/2^-$ and $3/2^-$  states of the same structure.

%
\begin{sidewaystable}
     \renewcommand{\arraystretch}{1.2}
\centering
\caption{Dimensionless coupling constants of the $(I=0, J^P=1/2^-)$
  poles found in this work. } \label{tab:cou12}
\begin{tabular}{ccccc ccccc}
\hline\hline
  & $\eta_c \Lambda$ & $J/\psi \Lambda$ &  $\bar D \Xi_c$ &  $\bar D_s \Lambda_c$ &  $\bar D \Xi'_c$ &  $\bar D^* \Xi_c$
  & $\bar D^*_s  \Lambda_c$ & $ \bar D^* \Xi'_c$  &  $\bar D^* \Xi^*_c$    \\
\hline
\multicolumn{2}{c}{$4276.59+i7.67$}  \\
\hline
$g_i$ & $0.17-i0.03$ & $0.29-i0.07$ & $\mathbf{2.93+i0.08}$ & $0.76+i0.31$ & $0.00+i0.01$ & $0.01+i0.02$ & $0.01+i0.04$ & $0.01-i0.02$ & $0.01-i0.03$  \\
$|g_i|$ & $0.17$ & $0.30$ & $\mathbf{2.93}$ & $0.82$ & $0.01$ & $0.02$ & $0.05$ & $0.02$ & $0.03$  \\
\hline
\multicolumn{2}{c}{$4429.84+i7.92$}  \\
\hline
$g_i$ & $0.29-i0.11$ & $0.17-i0.07$ & $0.00-i0.00$ & $0.00-i0.00$ & $0.15-i0.26$ & $\mathbf{2.78+i0.01}$ & $0.66+i0.32$ & $0.01+i0.05$ & $0.01+i0.03$  \\
$|g_i|$ & $0.31$ & $0.18$ & $0.00$ & $0.00$ & $0.30$ & $\mathbf{2.78}$ & $0.73$ & $0.05$ & $0.04$  \\
\hline
\multicolumn{2}{c}{$4436.70+i1.17$}  \\
\hline
$g_i$ & $0.24+i0.03$ & $0.14+0.01$ & $0.00-i0.00$ & $0.00-i0.00$ & $\mathbf{1.72-i0.04}$ & $0.22-i0.31$ & $0.06-i0.01$ & $0.01-i0.04$ & $0.01-i0.03$  \\
$|g_i|$ & $0.24$ & $0.14$ & $0.00$ & $0.00$ & $\mathbf{1.72}$ & $0.38$ & $0.07$ & $0.04$ & $0.03$  \\
\hline
\multicolumn{2}{c}{$4580.96+i2.44$}  \\
\hline
$g_i$ & $0.12-i0.00$ & $0.37-i0.04$ & $0.02-i0.01$ & $0.02-i0.01$ & $0.03-i0.00$ & $0.02-i0.02$ & $0.03-i0.02$ & $\mathbf{1.57-i0.17}$ & $0.00+i0.02$  \\
$|g_i|$ & $0.12$ & $0.37$ & $0.02$ & $0.02$ & $0.03$ & $0.03$ & $0.03$ & $\mathbf{1.58}$ & $0.02$  \\
\hline
\multicolumn{2}{c}{$4650.86+i2.59$}  \\
\hline
$g_i$ & $0.32-i0.05$ & $0.19-i0.03$ & $0.02-i0.01$ & $0.03-i0.02$ & $0.02-i0.00$ & $0.01-i0.01$ & $0.02-i0.01$ & $0.01-i0.00$ & $\mathbf{1.41-i0.23}$  \\
$|g_i|$ & $0.32$ & $0.19$ & $0.03$ & $0.04$ & $0.02$ & $0.02$ & $0.02$ & $0.02$ & $\mathbf{1.43}$  \\
\hline
\end{tabular}
\end{sidewaystable}
%

\begin{table}[ht]
     \renewcommand{\arraystretch}{1.2}
\centering
\caption{Same as Table~\ref{tab:cou12} for $J^P= 3/2^-$.} \label{tab:cou32}
\begin{tabular}{cccc ccc}
\hline\hline
 & $J/\psi \Lambda$ &  $\bar D^* \Xi_c$ &  $\bar D_s^* \Lambda_c$ 
&  $\bar D^* \Xi'_c$  &  $\bar D \Xi^*_c$  & $\bar D^* \Xi_c^*$  \\
\hline
\multicolumn{2}{c}{$4429.52+i7.67$}  \\
\hline
$g_i$ & $0.31-i0.10$ & $\mathbf{2.77-i0.02}$ & $0.67+i0.32$ & $0.00+i0.0.02$ & $0.00-i0.06$ & $0.00+i0.0.04$  \\
$|g_i|$ & $0.32$ & $\mathbf{2.77}$ & $0.74$ & $0.02$ & $0.06$ & $0.04$  \\
\hline
\multicolumn{2}{c}{$4506.99+i1.03$}  \\
\hline
$g_i$ & $0.27-i0.02$ & $0.02-i0.03$ & $0.02-i0.02$ & $0.00-i0.03$ & $\mathbf{1.56-i0.07}$ & $0.00-i0.05$  \\
$|g_i|$ & $0.27$ & $0.03$ & $0.03$ & $0.03$ & $\mathbf{1.56}$ & $0.05$   \\
\hline
\multicolumn{2}{c}{$4580.96+i0.34$}  \\
\hline
$g_i$ & $0.14-i0.01$ & $0.01-i0.01$ & $0.01-i0.01$ & $\mathbf{1.54-i0.02}$ & $0.02-i0.00$ & $0.00-i0.04$  \\
$|g_i|$ & $0.14$ & $0.01$ & $0.02$ & $\mathbf{1.54}$ & $0.02$ & $0.04$   \\
\hline
\multicolumn{2}{c}{$4650.58+i1.48$}  \\
\hline
$g_i$ & $0.29-i0.02$ & $0.02-i0.01$ & $0.03-i0.02$ & $0.03-i0.01$ & $0.03-i0.00$ & $\mathbf{1.40-i0.13}$  \\
$|g_i|$ & $0.29$ & $0.03$ & $0.03$ & $0.03$ & $0.03$ & $\mathbf{1.41}$   \\
\hline
\end{tabular}
\end{table}
One should remind the reader that there are suggestions of reactions
to see these states, like the $\Lambda_b \to J/\psi K^0 \Lambda$
reaction \cite{Lu:2016roh}, $\Lambda_b \to J/\psi \eta \Lambda$
\cite{Feijoo:2015kts}, $\Xi_b^- \to J/\psi K^- \Lambda$
\cite{Chen:2015sxa}. It should be noted that the $\Xi_b^- \to J/\psi
K^- \Lambda$ reaction has already been observed by the LHCb
collaboration \cite{Aaij:2017bef} and results of the LHCb Run-2 experiment
should be under present analysis, as anticipated in
Ref.~\cite{Aaij:2017bef}, searching for possible peaks in the $J/\psi
\Lambda$ mass distribution.

\section{Conclusions}

We have studied the interaction of coupled channels of meson-baryon
involving $c\bar{c}$ quarks, an $s$ quark and zero isospin. There are nine channels
coupling to $1/2^-$ in $S-$wave, six channels coupling to $3/2^-$ and
just one channel contributing to $5/2^-$. The interaction is
constructed implementing the symmetries of heavy quark spin physics
which provides a few independent matrix elements. The strength of
these terms is taken from an extension of the LHG approach to the
charm sector which turns out to be rather successful in the
description of the $\Omega_c$ states of LHCb \cite{Aaij:2017nav} in
Refs.~\cite{Debastiani:2017ewu,Montana:2017kjw}, and the recent
pentaquark states of Ref.~\cite{Aaij:2019vzc} in
Ref.~\cite{Xiao:2019aya}. This approach in the light sector leads to
the LO chiral Lagrangians. In the leading terms, obtained from the
exchange of light vectors, the $c$ quarks are mere spectators and
hence the interaction is independent of them and automatically
respects the rules of heavy quark spin symmetry. We obtain five
$1/2^-$ states, four $3/2^-$ states and one $5/2^-$ state.  We clearly
identify three near degenerate HQSS multiplets corresponding to states that couple most
strongly to $\bar{D}^*\Xi_c\ (1/2^-,\ 3/2^-)$,
$\bar{D}^*\Xi_c'\ (1/2^-,\ 3/2^-)$ and 
$\bar{D}^*\Xi_c^*\ (1/2^-,\ 3/2^-,\ 5/2^-)$. Other states appear with
just one spin, which are the states coupling mostly to
$\bar{D}\Xi_c\ (1/2^-)$, $\bar{D}\Xi_c'\ (1/2^-)$ and
$\bar{D}\Xi_c^*\ (3/2^-)$, though the mass difference between the
two latter states is very similar to that of the $\Xi_c^\prime$ and
$\Xi^*_c$ baryons. Indeed, these two dynamically generated states
form a further HQSS doublet broken by the $\Xi_c^\prime-\Xi^*_c$ mass
splitting.

The spectrum of states found in this work can be
easily understood from the symmetry remarks made at
the end of Subsec.~\ref{susec:theory}. We have found two distinct sets of
states. The first set (two $1/2^-$ resonances and a $3/2^-$ one)
results from the $\bar D^{(*)}\Xi_c-\bar D^{(*)}_s\Lambda_c$ coupled-channels
isoscalar interaction.  These resonances are significantly  broader than the others, 
with widths of the order of 15 MeV, and they decay mostly to
$D^{(*)}_s\Lambda_c$. Seen as $\bar D^{(*)}\Xi_c$ quasi-bound states, they
would be placed significantly below the corresponding thresholds,
around 50-60 MeV.

The second set is composed of quasi-bound  $\bar D^{(*)}\Xi_c^{\prime,
  *}$ isoscalar states, located around 5 to 10 MeV below their
thresholds. They are very narrow and would be the  SU(3)-partners of the
isospin 1/2 $\bar D^{(*)} \Sigma_c^{*}$ states obtained in
\cite{Xiao:2019aya} using the same formalism, that naturally
accommodated the $P_c(4440)$, $P_c(4457)$ and $P_c(4312)$ LHCb exotic states

The success in the description of the pentaquark states reported in
Ref.~\cite{Aaij:2019vzc} with the present approach, and the use here
of the same regulator used there for the loop functions, makes the
results obtained here rather credible, and it will be very interesting
to compare them with results of future experiments, in particular from
the study of the $\Xi_b^- \to J/\psi K^- \Lambda$ reaction that could
be the first one providing information of these states from the
analysis of the Run-2 experiments.

\section*{Acknowledgments}
This research  has been supported by the Spanish Ministerio de
Ciencia, Innovaci\'on  y Universidades and European FEDER funds under
Contracts FIS2017-84038-C2-1-P,  FIS2017-84038-C2-2-P and
SEV-2014-0398 and by the EU STRONG-2020 project under the program
H2020-INFRAIA-2018-1, grant agreement no. 824093.


\end{document}